\documentclass[aps,prl,showpacs,twocolumn]{revtex4}
\usepackage{epsfig}
\usepackage{amsfonts}

\begin{document}


\title{Soliton "molecules": Robust clusters of light bullets}


\author{Lucian-Cornel Crasovan$^*$, Yaroslav V. Kartashov$^\dag$, Dumitru Mihalache$^*$, Lluis Torner}
\affiliation{Institute of Photonic Sciences, Universitat
Politecnica de Catalunya, ES 08034 Barcelona, Spain}
\author{Yuri S. Kivshar}
\affiliation{Nonlinear Physics Group, Research School of Physical
Sciences and Engineering, the Australian National University,
Canberra, Australia}
\author{V\'{\i}ctor M. P\'erez-Garc\'{\i}a}
\affiliation{Departamento de Matem\'aticas, E.T.S.I.~Industriales,
Universidad de Castilla-La Mancha, 13071 Ciudad Real, Spain}


\begin{abstract}

{We show how to generate robust self-sustained clusters of soliton
bullets-spatiotemporal (optical or matter-wave) solitons. The
clusters carry an orbital angular momentum being supported by
competing nonlinearities. The "atoms" forming the "molecule" are
fully three-dimensional solitons linked via a staircase-like
macroscopic phase. Recent progress in generating
atomic-molecular coherent mixing in
Bose-Einstein condensates might open potential scenarios for the
experimental generation of these soliton molecules with matter-waves.}

\end{abstract}


\maketitle

Solitons - non-spreading, self-sustained wave packets - are at the core of
nonlinear science, thus they have been investigated and observed in a variety
of settings during the last two decades
\cite{Akhmediev}. Today, one of the most challenging open
frontiers of the field is the elucidation of complex soliton structures or
"soliton molecules" to be constructed from a number of "atoms", each being a
fundamental soliton. However, multi-soliton structures found so far
\cite{SoljacicPRL98}-\cite{chi24} tend to self-destroy
through expansion or collapse, or at best exist as meta-stable states which
break apart by small perturbations. Here we reveal, for the first time to our
knowledge, a physical mechanism for generating clusters which are made of
stable fully three-dimensional light bullets that propagate stably over huge
distances even in the presence of random perturbations in the initial
conditions.  The core of our approach is the use of two-color parametric
solitons supported by competing nonlinearities \cite{william,Buryak02}, which
allow both, to generate stable fully three-dimensional solitons and to reduce
the soliton-soliton interactions and enhancing the clusters robustness. The
clusters are thus multicolored, carry orbital angular momentum, and are linked
via a staircase-like macroscopic phase distribution.  We present the analysis
for optical spatiotemporal solitons, but our findings are intended to stimulate
further theoretical and experimental research in the case of matter-waves in Bose-Einstein
condensates \cite{BEC1}-\cite{BEC4}.

Spatiotemporal optical solitons, the so-called "light bullets"
(LBs), are self-sustained objects localized in all spatial
dimensions and in time \cite{LB1}-\cite{tandem} (for a recent
overview see Ref.\cite{wise}). They result from the simultaneous
balance of diffraction and dispersion by the medium nonlinearity,
and a two-dimensional version has recently been generated in
quadratic nonlinear media \cite{Liu}. On one hand, spatiotemporal
solitons are challenging objects for fundamental research, as
examples of stable localized objects in three-dimensional
nonlinear fields are rare in physics. On the other hand,
spatiotemporal solitons hold promise for potential applications in
future ultrafast all-optical processing devices
\cite{Snyder}-\cite{liu2}, where each soliton represents a bit of
information and should be employed for digital operations.
Multi-channel all-optical soliton networks have been proposed
based on the concept of soliton clusters \cite{yuriopn}, the
structures carrying many interacting individual solitons, recently
introduced for two-dimensional solitons in saturable nonlinear
media \cite{DesyatnikovPRL02}.

Soliton clusters can be viewed as a nontrivial generalization of "spinning"
solitons (or doughnut-like vortices)
\cite{chi3}-\cite{MihalachePRL02} and necklace-ring beams
\cite{SoljacicPRL98}-\cite{DesyatnikovPRL01}, and they also appear
in the study of active nonlinear systems such as externally driven optical
cavities \cite{Vladimirov},\cite{Skryabin}. But the soliton clusters
investigated so far tend to be unstable or meta-stable under the action of
small perturbations. We have recently shown in the case of two-dimensional
spatial solitons, that the competition between quadratic and cubic
nonlinearities reduces the strength of the soliton-soliton interactions, thus
making spatial soliton clusters more robust under propagation \cite{yaros2}.
Here, we consider for the first time the case of clusters made of fully
three-dimensional light bullets, and show that they propagate stably over huge
distances even in the presence of random perturbations.

We consider the propagation of two-color (fundamental wave and second harmonic)
LB "molecules" (see the sketch in Fig. 1) in a bulk dispersive medium with
competing quadratic and cubic (Kerr) self-defocusing nonlinearities.
Under suitable conditions, the interaction between a fundamental frequency (FF)
signal and its second harmonic (SH), in the presence
of the self-defocusing cubic nonlinearity, dispersion and diffraction in the
(3+1)-dimensional geometry, can be described by the reduced model
\cite{Menyuk}-\cite{Bang}
\begin{eqnarray}
i\frac{\partial u}{\partial Z}+&&\frac{1}{2}\left( \frac{\partial
^{2}u} {\partial X^{2}}+\frac{\partial ^{2}u}{\partial
Y^{2}}+\frac{\partial ^{2}u}
{\partial T^{2}}\right) +u^{\ast }\,v\, \nonumber \\
&&-\alpha (|u|^{2}+2|v|^{2})u\,=0,  \nonumber \\
i\frac{\partial v}{\partial Z}+&&\frac{1}{4}\left( \frac{\partial
^{2}v} {\partial X^{2}}+\frac{\partial ^{2}v}{\partial
Y^{2}}+\sigma \frac{\partial ^{2}v}{\partial T^{2}}\right) -\beta
v+u^{2}\,  \nonumber \\
&&-2\alpha (2|u|^{2}+|v|^{2})v\,=0. \label{scaled}
\end{eqnarray}
Here, $T$, $X$, $Y$ and $Z$ are the normalized reduced time,
transverse spatial coordinates, and propagation distance, $u$ and
$v$ are envelopes of the FF and SH fields, $\alpha$  measures the
strength of the defocusing cubic nonlinearity, and $\beta$  is a
phase mismatch between the FF and SH waves. Here $^\ast$  stands
for the complex conjugate of a complex field. Equations
(\ref{scaled}) assume different group-velocity dispersion
coefficients at the two frequencies, $\sigma$ being their ratio,
and assumes that the temporal group-velocity mismatch between them
has been compensated. Notice that Eqs.(1) correspond to the simplest
model of light propagation in media with competing nonlinearities
(e.g., it assumes a non-critical, type I, oo or ee wave interaction).
In practice, the strength of each of the possible cross-phase-modulations
depends critically on the crystalline symmetry of the particular material
employed through the polarizations of the fields involved, hence the actual
value of the relevant elements of the nonlinear susceptibility tensor.
However, Eqs. (1) are  expected to capture the essential
physics behind the soliton cluster evolution.

The interaction Hamiltonian of the system is:
\begin{eqnarray}
H &=&\frac{1}{2}\int \int \int \left\{
(|u_{X}|^{2}+|u_{Y}|^{2}+|u_{T}|^{2})+ \right. \nonumber  \\
&&\frac{1}{4}(|v_{X}|^{2}+|v_{Y}|^{2}+ \sigma |v_{T}|^{2})
+ \beta |v|^{2}-(u^{\ast 2}v+u^{2}v^{\ast }) \nonumber \\
&& \left. +\alpha (|u|^{4}+4|u|^{2}|v|^{2}+|v|^{4}) \right\}
dXdYdT \label{inv2}
\end{eqnarray}
is a conserved quantity during evolution. Its absolute and local
minima correspond to stable and metastable configurations,
respectively.

Circular light-bullet necklaces were constructed as superposition
of $N$ fundamental spatiotemporal solitons with different phases
such that the overall phase jump around the core is a multiple of
$2\pi$ (see Fig. 1). We thus have:
\begin{eqnarray}
u(Z=0)&=&\sum_{n=1}^{N}u_0(\vec{r}-\vec{r}_n)e^{i\phi_n}~,
\nonumber
\\ v(Z=0)&=&\sum_{n=1}^{N}v_0(\vec{r}-\vec{r}_n)e^{2i\phi_n}~,
\label{cluster}
\end{eqnarray}
where $u_0$ , $v_0$  are the fundamental solitons at both
frequencies, $\vec{r}_n$ are the soliton locations, whereas the
soliton phases at those points are $\phi_n=2n\pi M/N$ and
$2\phi_n$ , respectively. Here $M$ determines the full phase twist
around the cluster and plays the role of a topological charge
("spin"). We have considered circular soliton arrays, i.e. equally
spaced "atoms" displaced on a circle of radius $R_0$ . First, by
appropriate numerical techniques (a standard band-matrix algorithm
to deal with the resulting two-point boundary-value problem) we
have found the families of stationary solutions to Eqs. (1) - i.e.
the fundamental (non-spinning) three-dimensional spatiotemporal
solitons ($u_0$ ,$v_0$). In fact, the stationary three-dimensional
parametric soliton can be well approximated by a super-Gaussian
"ansatz" with suitable chosen amplitudes and widths for both the
FF and SH fields.

The parameters that play an important role in the dynamics of the
LB "molecules" are the necklace topological charge $M$, the number
of "pearls"  $N$ forming the cluster, the initial radius of the
necklace $R_0$, the energy $E_{LB}$ of each constituent soliton,
the wave-vector mismatch $\beta$ and the strength of the
defocusing cubic nonlinearity $\alpha$. In almost all of our
calculations we have considered the phase-matching of the
interacting waves, taking thus $\beta=0$. We have also set
$\sigma=1$, assuming equal dispersions at both frequencies, and
$\alpha=0.2$ as the dynamical equations possess scaling properties
with respect to $\alpha$. By increasing the strength of the
defocusing cubic nonlinearity one will slow down the interaction
between the constituent "atoms". Taking into account that the
medium with competing nonlinearities supports stable
spatiotemporal vortices (vortex tori) with unit topological charge
when their energy exceeds a threshold \cite{MihalachePRE02}, we
have studied in detail the dynamics of soliton "molecules" which
have the total energy exceeding the corresponding stability
threshold energy of the vortex soliton. Because the energy
threshold for the existence of a stable vortex torus at
$\alpha=0.2$ is $E_{th} \approx 9120$ , we have considered here
clusters with $N=5$ and $N=6$ solitons, each constituent having
the energy $E_{LB}=2100$ , whereas for the cluster with $N=4$
"atoms", the individual energy $E_{LB}=2824$ was correspondingly
higher.

Firstly, we have studied the dependence of the cluster interaction
Hamiltonian (or equivalently, the effective potential, defined as
$H(R_0)/H(\infty)$) on the initial radius $R_0$ and on the
necklace charge $M$. This quantity gives important hints when
looking for soliton bound states (see, e.g. Ref. \cite{DesyatnikovPRL02}, \cite{Malomed}
for a detailed analysis). While the interaction Hamiltonian for
the $N=4$ clusters does not posses any minima whatever the
topological charge is (see Fig. 2(a)), for $N=5$ and $N=6$, local
minima of the Hamiltonian are present for charge $M=1$. For $N=5$
the minimum is at $R_0=13.5$, whereas for $N=6$ the minimum is at
$R_0=12$. In our simulations we have added normally distributed
noise with zero mean and variance $\sigma_{noise}=0.1$ to the
input "molecules". Keeping $M=1$, we have varied the initial
cluster radius $R_0$ around the minimum value given by the
effective potential approach and have found a range of optimal
values of the input radii that minimize the mean radius
oscillations of the soliton cluster. For $N=6$ the value $R_0=12$
lies in the optimal radius interval, whereas for $N=5$, the value
$R_0=16$ assures small oscillations of the mean radius.

In order to check the predictions given by the study of the
effective potential, we have numerically solved Eqs.
(\ref{scaled}) by using a finite-difference scheme based on a
Cranck-Nicholson time discretization followed by a Newton-Picard
iterative technique and the Gauss-Seidel method for solving the
obtained system of equations. Transparent boundary conditions
allowing the radiation to escape from the computation window have
been implemented. We have monitorized the evolution of the mean
radius of the cluster defined as:
\begin{equation}
R(Z)~= \frac{1}{E}~\int \int \int (X^{2} + Y^{2} + T^{2})^{1/2}
(\left| u\right| ^{2}+\left| v\right| ^{2})dXdYdT , \label{radius}
\end{equation}
where $E=\int \int \int (\left| u\right| ^{2}+\left| v\right|
^{2})dXdYdT$ is the total energy. If the initial radius $R_0$ of
the cluster is large, then the mean radius $R(0)$ at the entrance
of the nonlinear medium amounts to $R(0) \approx R_0$.

The evolution of clusters with $N=4$ ($R_0=12$) and $N=5$
($R_0=16$) constituents is quite robust as shown in Fig. 3. The
"molecules" undergo rotation and clean up the initial noise in the
first stages of propagation. Our estimations for the angular
velocity $\omega$ of the soliton clusters end up with
$\omega=0.0027$ for the $N=4$ cluster shown in Fig. 3 and
$\omega=0.0014$ for the $N=5$ one. Thus, cluster rotations are
observable after large propagation distances. Only after thousands
of diffraction lengths a quasi-periodic shrinking and expansion
followed by a decay into several unequal fragments is observed as
seen in Fig. 4. Our soliton clusters are much more robust than the
LB clusters in quadratic and cubic saturable materials that
survive only a few diffraction lengths in the presence of initial
random noise.

The simulations with other necklace charges ($M=0$, $M=2$ and
$M=3$) for clusters composed of $N=5$ ($R_0=16$) and $N=6$
($R_0=12$), show that the LBs forming the $M=0$ "molecule" fuse in
100 - 150 propagation units, whereas the soliton clusters with net
charges $M=2$ or $M=3$ expand indefinitely. Detailed simulations
performed for the $N=6$-light bullet clusters with $M=2$ show
that, by varying the initial cluster radius, the clusters formed
with overlapping solitons ($10<R_0<20$) expand rapidly whereas the
clusters built with well separated LB´s ($R_0>22$) have a moderate
mean radius variation for a propagation distance over 600
diffraction lengths. Notice that for a typical diffraction length of a few mm,
this corresponds to several meters, orders of magnitude larger than the feasible
crystal lengths. Similar results were obtained for the non
phase-matching case ($\beta \ne 0$).

We have also studied the influence of the initial phase
distribution on the cluster dynamics by  simulating the evolution
of two configurations with identical intensity distributions but
different phases.

The first one, build as per Eq. 3, having a staircase-like phase,
destroys finally, after thousands of diffraction lengths, by
splitting into two spatiotemporal solitons (Fig. 5(a)-(d)), while
the second one, having a ramp-like phase mask (see the inset of
Fig. 5(e)), develops into a vortex torus (Fig. 5(e)-(h)). Thus, we
arrive at the conclusion that the key factor that impede the LB
"molecule" with a staircase-like macroscopic phase to excite a
vortex soliton is the sequence of the phase edge-dislocations (see
the inset in Fig. 5(a)) existing between the neighboring solitons
which form the cluster.

In summary, we have revealed a key physical mechanism for creating
truly three-dimensional light bullet clusters which survive under
random perturbations of the initial conditions. We have generated
such structures numerically for a nonlinear optical medium with
competing quadratic and cubic nonlinearities. The experimental
demonstration of the concept with light waves faces many important
challenges, including the generation of single light bullet. This
goal requires the elucidation of a material setting with high
quadratic nonlinearity,  suitable group-velocity-dispersions and
low one-photon and two-photon absorption at both FF and SH
wavelengths, as well as small group-velocity-dispersion, together
with adequate cubic nonlinearities.  This is a formidable task,
thus progress is being made slowly. In this context we would like
to mention that it was shown recently that the strength of the
cubic nonlinearity can be tuned by means of optical rectification
\cite{Torres} even though at present the technique has been
developed only for one-dimensional beams.

However, although we showed the concept in the case of light
waves, our study is important to other fields such as the physics
of hybrid atomic-molecular Bose-Einstein condensates
\cite{Julienne}-\cite{Donley}.  Indeed, recent experiments
demonstrated coherent mixing of atomic-molecular condensates
\cite{Donley} which under suitable conditions should be
approximately described by coupled equations for the macroscopic
wave functions similar to Eqs.  (\ref{scaled})
\cite{Julienne}-\cite{Kokkelmans}. Taking into account that to
date the experimental observations of bright solitons in
condensates are restricted to quasi-one dimensional geometries
\cite{BEC1},\cite{BEC2}, the matter-wave analogue of our light
bullet clusters would correspond to clusters of condensate drops
existing without a trap.

{\bf Acknowledgements} This work has been supported by the Generalitat de
Catalunya and by the Spanish Government under contracts TIC2000-1010 and
BFM2002-2861.  Support from NATO (L.-C.C.) and IBERDROLA S. A., Spain (D.M.) is
acknowledged.

$^*$Also with: Department of Theoretical Physics, Institute of
Atomic Physics, PO Box MG-6, Bucharest, Romania; $^\dag$Permanent
address: Physics Department, M. V. Lomonosov Moscow State
University, Vorobiovy gory, 119899 Moscow, Russia.




\newpage

{\bf FIGURE CAPTIONS}

\vspace{2cm}

\noindent
{\bf Fig. 1} Cluster composed of six spatiotemporal
two-color solitons. The topological charge $M$ of the soliton
cluster is equal to one. (a) The fundamental frequency field and
(b) the second harmonic field. (c) The phase distribution at
fundamental frequency and (d) the phase distribution at the second
harmonic.
\vspace{0.2cm}

\noindent
{\bf Fig. 2} Effective interaction potential versus
initial cluster radius for (a) $N=4$, (b) $N=5$ and (c) $N=6$
soliton clusters for different net topological charges. Typical
oscillations of the mean cluster radius of solitons clusters with
"spin" $M=1$ for (d) $N=4$, $R_0=12$, $E_{LB}=2824$, (e) $N=5$,
$R_0=16$, $E_{LB}=2100$ and (f) $N=6$, $R_0=12$ and $E_{LB}=2100$.
\vspace{0.2cm}

\noindent
{\bf Fig. 3} Stable evolution of soliton clusters with
$M=1$ under superimposed input random noise. Shown are the contour
plots for the $N=4$ cluster: (a), $Z=0$; (b), $Z=25$; (c), $Z=50$
and the contour plots for the $N=5$ cluster: (d), $Z=0$; (e),
$Z=25$; (f), $Z=50$. Only the $(X,Y)$ slices at $T=0$ of the
fundamental frequency component are shown; the second harmonic
field exhibits a similar behavior. The other parameters are the
same as in Fig. 2(d) for the $N=4$ cluster and as in Fig. 2(e) for
the $N=5$ one. \vspace{0.2cm}

\noindent
{\bf Fig. 4} Cluster evolution over long distances and
the onset of symmetry breaking instability. Shown are the
isosurfaces $|u|=1.1$ for the $N=4$ (a)-(d) and the $N=5$ cluster
(e)-(h). The parameters are the same as in Figs. 2(d) and 2(e).
\vspace{0.2cm}

\noindent
{\bf Fig. 5} Comparative evolution of two clusters with
identical intensity distributions but different phase masks. The
net topological charge is the same ($M=1$) in both situations. Top
panels: evolution of a six-soliton "molecule" with a step-like
phase distribution; bottom panels: evolution of a six-soliton
"molecule" with a ramp-like phase distribution. Shown are the
isosurfaces $|u|=1.1$. The insets in panels (a) and (e) show the
initial phase mask.

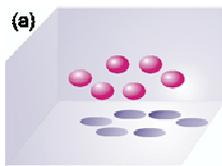
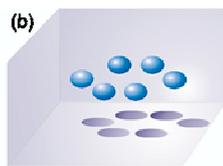
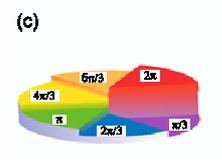
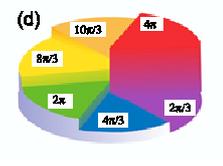
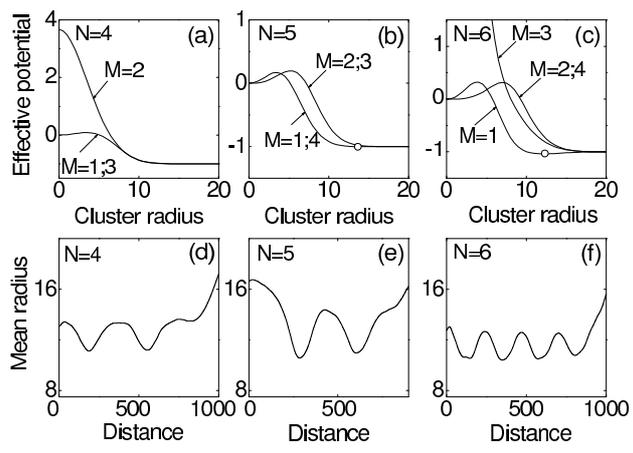
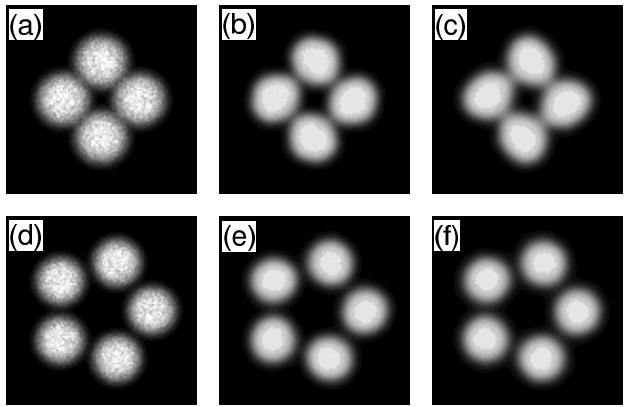
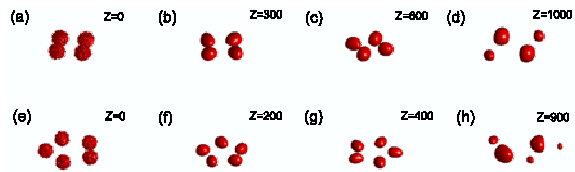
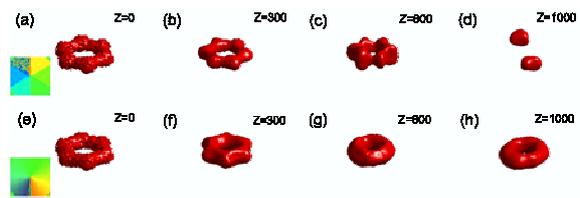
\end{document}